\documentstyle[psfig,epsf]{mn}

\title[Observations of KPD2109+4401]
{Resolving the pulsations of the subdwarf B star KPD~2109+4401}

\author[A.-Y. Zhou et al.]
{A.-Y. Zhou,$^{1,2}$ M. D. Reed,$^1$\thanks{E-mail:
MikeReed@missouristate.edu}
S. Harms,$^1$  D. M. Terndrup,$^3$ D. An,$^3$ S. Zola,$^{4,5}$\cr
K. D. Gazeas,$^6$ P. G. Niarchos,$^6$ W. Ogloza,$^4$ A. Baran$^{4,7}$ 
and G. W. Wolf$^1$ \\
$^1$Department of Physics, Astronomy and Materials Science, 
Missouri State University, 901 S. National, Springfield, MO 65804 USA \\
$^2$National Astronomical Observatories, Chinese Academy of Sciences, 
Beijing 100012, China\\
$^3$Department of Astronomy, The Ohio State University,
140 W. 18th Ave., Columbus, OH 43210 USA\\
$^4$Mt. Suhora Observatory of the Pedagogical University, ul. 
Podchor\c{a}zych 2, PL-30-084 Cracow, Poland \\
$^5$Astronomical Observatory, Jagiellonian 
University, ul. Orla 171, 30-244 Cracow\\
$^6$Department of Astrophysics, Astronomy and
Mechanics, Faculty of Physics, University of Athens, GR 157 84,
Zografos, Athens, Greece\\
$^7$Torun Centre for Astronomy, ul. Gagarina 11, 87-100 Torun, Poland\\
}

\date{Accepted     
      Received }

\begin{document}

\maketitle

\begin{abstract}
%
We present the results of extensive time series photometry of
the pulsating subdwarf B star KPD 2109+4401.
Our data set consists of 29 data runs with
a total length of 182.6 hours over 31 days, 
collected at five observatories in 2004.
These data are comprised of high signal-to-noise 
observations acquired with larger telescopes 
and wider time coverage observations obtained with smaller telescopes.
They are sufficient to resolve the pulsation structure to 0.4~$\mu$Hz
and are the most extensive data set for this star to date.
With these data, we identify eight
pulsation frequencies extending from 4701 to 5481\,$\mu$Hz,
corresponding to periods of 182 to 213\,s.
The pulsation frequencies and their amplitudes are examined over several
time-scales with some frequencies showing
amplitude variability.
\end{abstract}


\section{Introduction}
Subdwarf B (sdB) stars are low-mass ($\sim$0.5\,M$_{\odot}$) core 
helium-burning
horizontal branch stars with very thin outer hydrogen layers,
making them quite hot; 
they likely proceed directly to the white dwarf cooling track without
reaching the asymptotic giant branch after core helium exhaustion
(Heber 1984; Saffer et al. 1994).
In recent years, over 30 sdB stars have been identified as
multimode pulsators, with typical pulsation periods of 100--250\,s
and  amplitudes generally less than a few hundredths of a magnitude.
Officially designated V361 Hya stars, they are also
commonly known as EC~14026 stars after the prototype (EC~14026-2647)
and referred to as sdBV stars following the pulsating white dwarf convention 
(DOV, DBV, DAV). Recent reviews of pulsating sdB stars have been given by
Charpinet, Fontaine \& Brassard (2001; pulsation theory) and 
Kilkenny (2002; observation).

Asteroseismology has successfully been applied to some other classes
of variable stars in order to discern their interior conditions 
(Winget et al. 1991; Kanaan et al. 2005; Mukadam et al. 2003; among others) and
in time, it is hoped the same can be accomplished for sdB stars.
To do so, the pulsation frequencies (periods) must first be resolved.
Variable star discovery surveys seldom resolve or detect the complete set of
pulsations. Multisite campaigns, because of the complexity
of organization, have only observed a few sdB pulsators.
Our program is to resolve poorly-studied sdB pulsators
from single-site data, supplemented with a small amount of 
multi-site data. This method has proven useful
for the sdBV star Feige 48 (Reed et al. 2004).
By combining limited (5--10 days) amounts of data from larger telescopes with
a longer timebase (some 20+ days) of data from smaller (0.4--1.0m) telescopes,
it is possible to effectively resolve the pulsation spectrum of an sdBV star.
Our high signal-to-noise (S/N) data from larger telescopes can detect
pulsation amplitudes as low as 0.2 milli-modulation amplitudes (mma, 
equivalent to 0.02\%), insuring that
we do not miss low amplitude modes and that the resolution of our complete data
set is 0.4~$\mu$Hz.

This paper reports the results of our observations on the sdBV star
KPD~2109+4401 (hereafter KPD~2109). KPD~2109 was discovered to be a
pulsator nearly simultaneously by Koen (1998; hereafter K98) and
Bill\'{e}res et al. (1998), both of whom detected five frequencies. 
Additional studies of KPD~2109 include a time-series spectroscopic
study (Jeffery \& Pollacco 2000) and multicolour observations using
ultracam on the WHT 4.2-m telescope (Jeffery et al. 2004). However, none
of these relatively short duration
observations have likely resolved the pulsation structure (in
fact, K98 claim not to have), prompting us to choose KPD~2109 for
follow-up observations.

\section{The observations}
\begin{figure*} \centering
\centerline{\psfig{figure=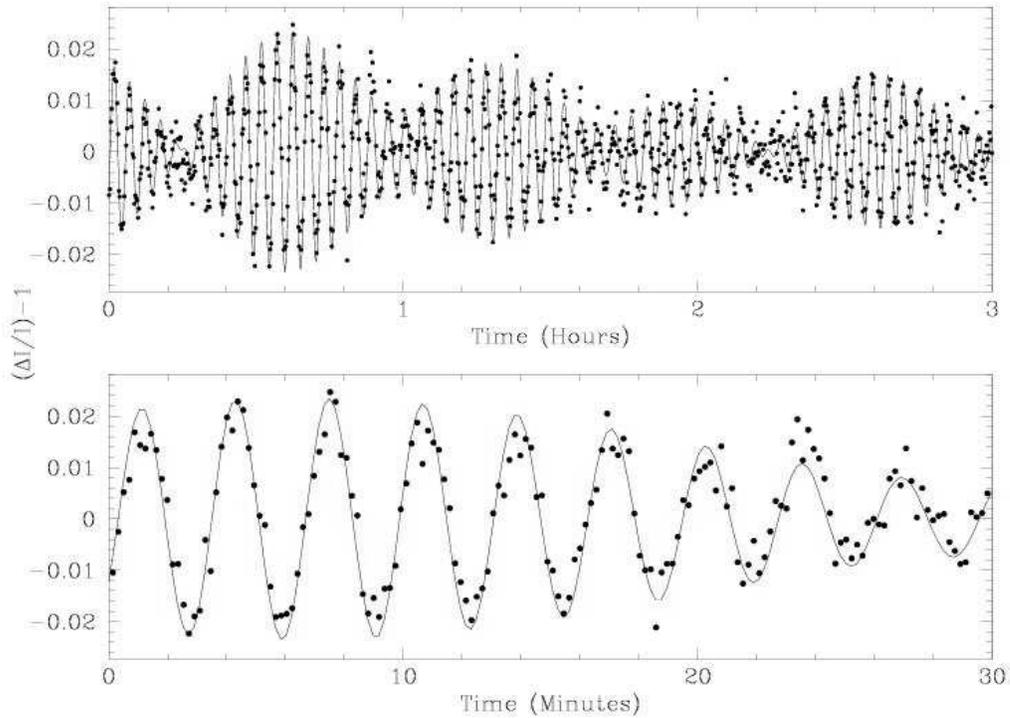,width=5.5in}}
\caption{Lightcurves for KPD~2109 obtained at MDM observatory, plotted
on two different time scales. The top panel shows 3 hours of data in which
the beating is obvious. The bottom panel shows 30 minutes of data to
emphasize individual pulsations. Solid line is the least-squares 
solution.}
\label{fig01}
\end{figure*}

Data were collected at five observatories
during September and October of 2004 with CCD photometers and the specifics
of each run are provided in Table~\ref{tab01}.
Data were obtained at the McDonald observatory 2.1~m telescope
and the MDM observatory 1.3~m telescope using an Apogee Alta U47+ CCD camera.
This camera is connected via USB2.0 for high-speed readout and our 
binned ($2\times 2$) images had an average dead-time of one second.
Observations at Baker Observatory (0.4~m) were obtained with 
a Princeton Instruments
RS1340 CCD camera. We used a 601$\times$601 pixel subframe at $1\times 1$ 
binning and our average dead-time was one second.
Data from the Observatory of the University
of Athens (0.4~m) were obtained with
an SBIG ST-8 CCD camera with an average dead-time of six
seconds. Photometry from
Mt. Suhora Astronomical Observatory, Poland (0.6~m) were obtained 
with an SBIG ST-10 XME CCD camera with an 
average dead-time of three seconds.
The observations at MDM, McDonald, and Baker observatories used red cut-off
filters (BG38, BG40, and BG40, respectively), Suhora observatory observations
used a Johnson B filter, and data from Athens used no filter. As pulsations
from sdB stars have little dependence in the visual between these various
filters, and no phase dependence (K98),  
mixing these data is not seen as a problem. The same is true for timing
from the various observatories: NTP was used at Athens, McDonald, Baker, and
MDM observatories and a GPS clock at Suhora observatory. Both of these
issues (particularly timing)
would also appear as artifacts in our analyses in \S 3 and 4.3.
 A portion of
data from MDM observatory is shown in Figure~\ref{fig01} on two different
scales. The top panel has several hours of data, 
showing the obvious beating that occurs on multiple time scales which
indicates the multiperiodic nature of the pulsations. The bottom panel
spans less time to emphasize individual pulsations.

\begin{table*}
\centering
\caption{Observations of KPD~2109+4401 \label{tab01}}
\begin{tabular}{|lcrl|lcrl|} \hline
Run & Length &
Date & Observatory & Run & Length & Date & Observatory\\
 & (Hrs) & UT & & & (Hrs) & UT &  \\ \hline
grc040912 & 9.5 & 12 Sept. & Athens 0.4m &bak092804 & 8.4 & 28 Sept. & Baker 0.4m \\
grc040913 & 3.6 & 13 Sept. & Athens 0.4m &bak092904 & 0.5 & 29 Sept. & Baker 0.4m \\
grc040914 & 5.6 & 14 Sept. & Athens 0.4m &bak093004 & 7.3 & 30 Sept. & Baker 0.4m \\
mdr269 & 8.4 & 14 Sept. & McDonald 2.1m  & bak100304 & 7.9 & 3 Oct. & Baker 0.4m\\
grc040915 & 3.5 & 15 Sept. & Athens 0.4m & mdr274 & 5.6 & 5 Oct. & MDM 1.3m \\
mdr270 & 8.8 & 15 Sept. & McDonald 2.1m  & mdr276 & 6.3 & 6 Oct. & MDM 1.3m\\ 
mdr271 & 8.7 & 16 Sept. & McDonald 2.1m  & mdr278 & 6.5 & 7 Oct. & MDM 1.3m\\ 
mdr272 & 8.3 & 17 Sept. & McDonald 2.1m  & mdr280 & 6.8 & 8 Oct. & MDM 1.3m\\ 
suh040918 & 6.8 & 18 Sept. & Suhora 0.6m  & mdr282 & 6.5 & 9 Oct. & MDM 1.3m\\ 
mdr273 & 3.1 & 19 Sept. & Baker 0.4m  & mdr284 & 2.0 & 10 Oct. & MDM 1.3m\\ 
bak092104 & 8.5 & 21 Sept. & Baker 0.4m  & mdr287 & 6.5 & 11 Oct. & MDM 1.3m\\ 
bak092204 & 8.3 & 22 Sept. & Baker 0.4m  & mdr289 & 1.7 & 12 Oct. & MDM 1.3m\\ 
bak092304 & 8.5 & 23 Sept. & Baker 0.4m  & mdr292 & 6.3 & 13 Oct. & MDM 1.3m\\ 
bak092604 & 8.4 & 26 Sept. & Baker 0.4m  & mdr294 & 1.8 & 14 Oct. & MDM 1.3m\\ 
bak092704 & 8.5 & 27 Sept. & Baker 0.4m  &  & &  & \\ \hline
\end{tabular}
\end{table*}

The standard procedures of image reduction, including bias subtraction,
dark current and flat field correction, were followed using IRAF packages.
Differential magnitudes were extracted from the calibrated
images using {\sc momf} (Kjeldsen \& Frandsen 1992).
Observations acquired at MDM observatory
generally had photometric conditions while others were through light
clouds or transparency variations. 
As sdB stars are substantially hotter, and thus bluer, than typical field
stars, differential light curves, even using an ensemble of comparison stars,
are not flat due to differential atmospheric and colour extinctions.
A low-order polynomial was fit to remove these trends from the data on a
night-by-night basis.

\section{The frequency content of the observations}

\begin{figure*} \centering
\centerline{\psfig{figure=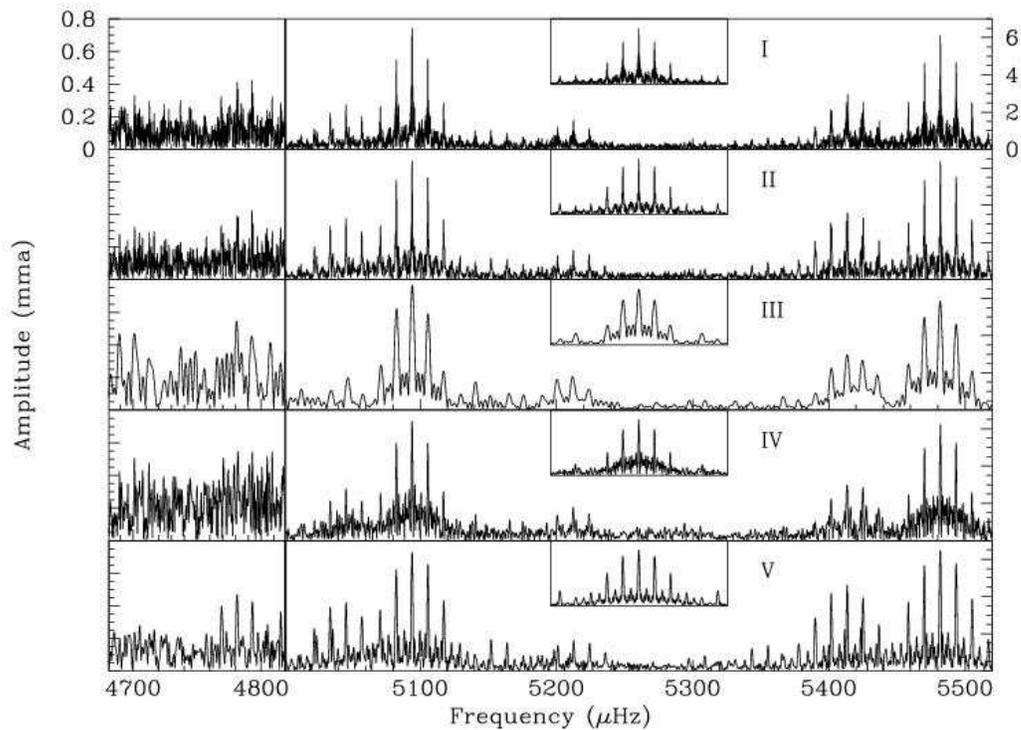,width=5.5in}}
\caption{Temporal spectra of KPD~2109 for
the groups in Table~\ref{tab02}. Note that only regions with
pulsations are included and that the scales for the left and right
panels are different. Insets are the spectral windows with horizontal scales
to match the right panels. }
\label{fig02}
\end{figure*}

Due to the length of our campaign we
grouped combinations of nightly runs into the sets given in
Table~\ref{tab02} for the convenience of analyses. Because the Athens
data was sufficiently noisier than other data, it was only included in 
Group I. The last two
columns of Table~\ref{tab02} give our frequency resolution and 
detection limit as calculated using the methods of Breger et al. (1994). The
detection limit serves as the limit below which pulsations cannot
be considered significant compared to the noise. We analyzed these
data in the usual manner (Kilkenny et al. 1999) which involves
examining subgroups of data for consistency. This also allowed us to examine
variations in amplitude over the course of our run and look for systematic
differences in phase or timing between observatories (none were detected).
Temporal spectra and window functions
 for the groups are plotted in Figure~\ref{fig02}. A window function is a
single, noise-free sine wave (of arbitrary amplitude) 
sampled at the same times as the data. The central peak of the window 
is the input frequency with other peaks indicating the
aliasing pattern of the data. Frequencies,
amplitudes and phases were determined by simultaneously fitting a nonlinear
least-squares solution to the data. Our solutions for the frequencies
for all groups as well as those discovered by K98
are provided in Table~\ref{tab03}.
We also calculated a 
noise-weighted FT for the Group I data, but do not show it as there was no
improvement.
This indicates that the majority of our data are quite good quality, with
the noisy runs not contributing sufficiently to increase the noise level.

\begin{table}
\centering
\caption{Subgroups used in pulsation analysis. \label{tab02}}
\begin{tabular}{lcccc}\hline
Group & Dates & Telescope(s) & Res. & Limit \\ 
      &   (2004)    &              & ($\mu$Hz)  & (mma) \\ \hline
I & 12 Sept. - 14 Oct. & All & 0.4 & 0.38 \\
II & 19 Sept. - 14 Oct. & Bak + MDM & 0.5 & 0.38 \\
III & 14 Sept. - 17 Sept. & McD+Suh & 3.5 & 0.67 \\
IV & 19 Sept. - 3 Oct. & Bak & 0.8 & 0.83 \\
V & 5 Oct. - 14 Oct. & MDM & 1.3 & 0.29 \\ \hline
\end{tabular}
\end{table}

\begin{table*}
\centering
\caption{Comparison of pulsations detected in the various groups. 
Formal least-squares errors are provided in parentheses. \label{tab03}}
\begin{tabular}{lllllll} \hline
Group & $f1$ & $f2$  & $f3$  & 
$f4$ & $f5$ & $f6$\\ \hline
I & 5481.819(3) & 5413.819(6) & 5413.002(7) & 5212.604(10) &
5093.981(2) & 5045.466(8) \\
II & 5481.819(2) & 5413.818(6) & 5413.000(7) &  5212.592(10) &
5093.974(2) & 5045.427(22) \\
III & 5481.804(41) & 5413.269(80)$^{\dag}$ & -- & 5212.497(155)
& 5094.189(37) & 5033.632(271)$^{\star}$ \\
IV & 5481.831(14) & 5413.460(31)$^{\dag}$ & -- & 5212.430(47) &
5093.986(15) & 5045.505(45)\\
V & 5481.811(11) & 5413.484(17)$^{\dag}$ & -- & 5212.629(41) &
5093.964(12) & 5045.504(20) \\
K98 & 5481.76 & 5413.62 & 5412.69 & 5212.42 & 5093.93 & 
5045.54 \\
\multicolumn{7}{l}{$^{\dag}$ The doublet is unresolved in these Groups.}\\
\multicolumn{7}{l}{$^{\star}$ Indicates modes offset by approximately the daily alias (11.56~$\mu$Hz).} \\ \hline
\end{tabular}
\end{table*}

\begin{figure*}
\psfig{figure=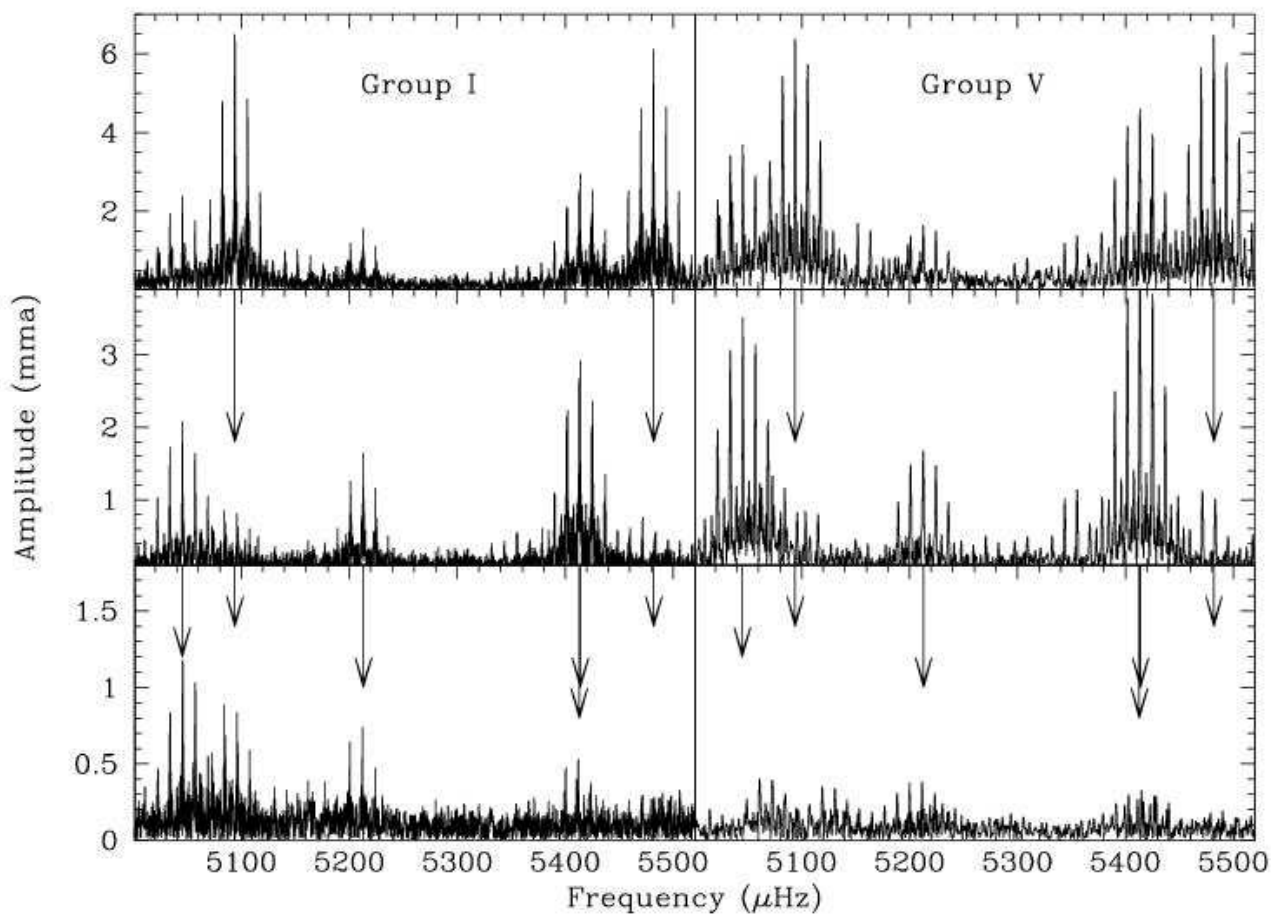,width=\textwidth}
\caption
{
Sequence of prewhitening for Groups I and V data. Top panels: Original
Fourier transforms. Middle panels: Prewhitened by the highest 2 peaks.
Bottom panels: Prewhitened by all 6 frequencies  in Table~\ref{tab03}.} 
\label{fig03}
\end{figure*}

\begin{figure*}
\caption
{Phases and amplitudes of the four well-separated frequencies over the 
course of our observations. Frequencies are indicated in each panel. Phases
are calculated as time of first maximum after JD=2453249.5 divided by the
period. Amplitudes from K98 and J04 are also provided (as filled squares and 
triangles, respectively).}
\psfig{figure=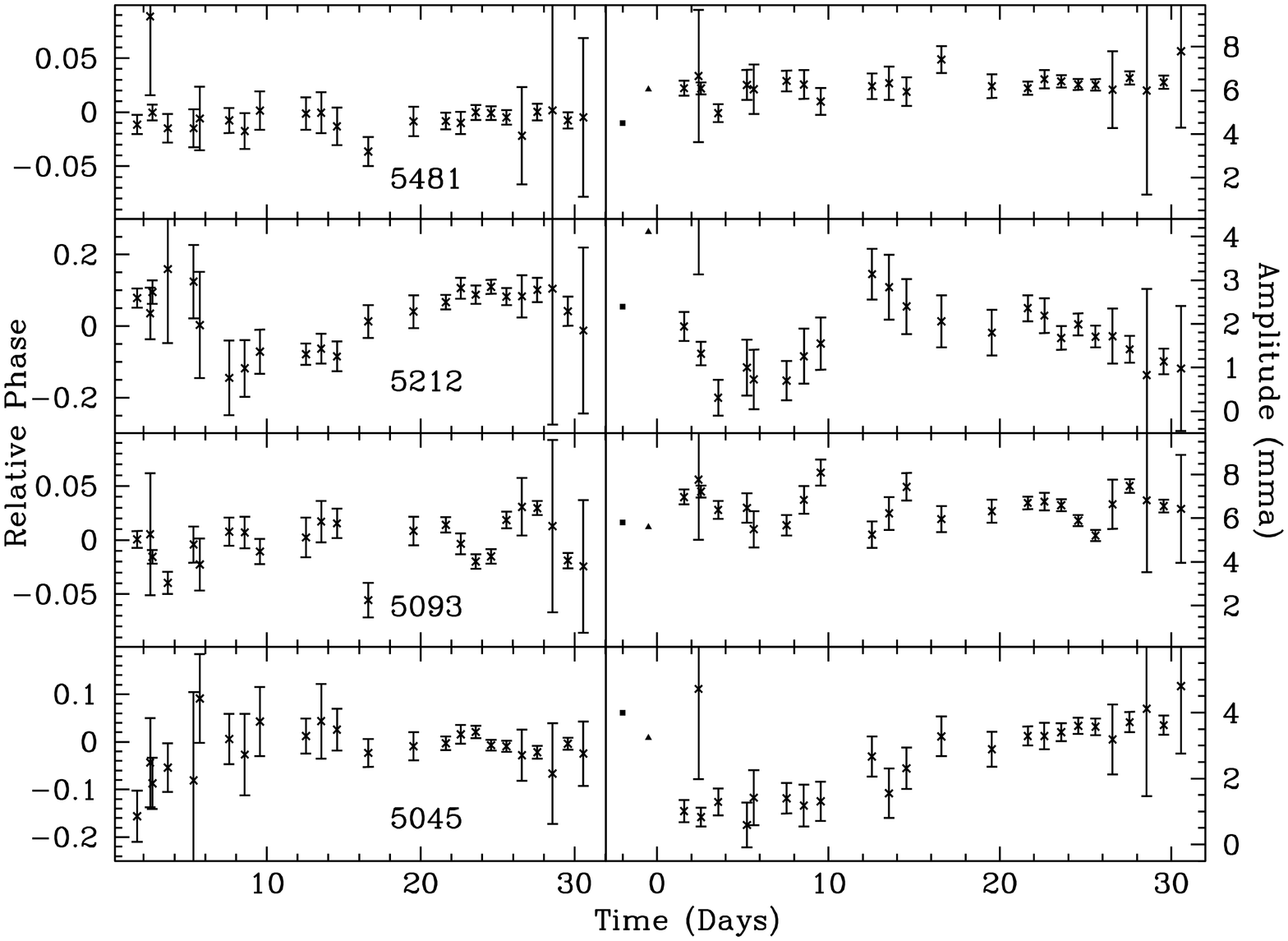,angle=-90,width=\textwidth}
\label{fig04}
\end{figure*}

The top panel of Figure~\ref{fig03} 
shows the original temporal spectrum for all the data
(Group I; left panels) as well as for just the MDM data (Group V; 
right panels). The middle panels show the residuals
after prewhitening by the highest two peaks  and the bottom panels are after
prewhitening by the highest six peaks.
Though some power remains in the Fourier transform (FT), and we could
continue prewhitening, the peaks are conspicuously close to previously
prewhitened frequencies and may 
be due to amplitude variation over the course of the
data run. To investigate this possibility,  Figure~\ref{fig04} shows 
the amplitudes of the four reasonably
separated modes over the course of our observations. (The doublet cannot
be separated in individual runs, so we do not fit it.) 
As anticipated, the frequencies with the largest residuals in 
Fig.~\ref{fig03} show the largest amount of variability in
their amplitudes (left panel) while the phases remain
within 20\% of their original value. This indicates that the amplitude
variability is intrinsic to the pulsations rather than due to beating 
between closely spaced modes. As prewhitening removes a constant-amplitude
sine curve from the data, it sometimes overestimates and
at other times underestimates the amplitudes. The net effect is that
prewhitening does not remove all of the power, leaving the residuals in
the bottom panel of Figure~\ref{fig03}. The right panel of that figure 
is just for the MDM
data and indicates that over the shorter time span, the amplitude variations
are less, so prewhitening does a better job and the residuals are 
correspondingly smaller.
This indicates that we have likely resolved the pulsation spectrum of
KPD~2109. Our least-squares  solution for the entire data set which
is provided in Table~\ref{tab04}.

\begin{table}
\centering
\caption{Non-linear least-squares solution to the entire data set. 
Formal least-squares errors are in parentheses.  \label{tab04}}
\begin{tabular}{llll}
Label & Period & Frequency & Amplitude  \\ 
 & (sec) & ($\mu$Hz) & (mma)  \\ \hline
$f1$ & 182.42120(8)  &  5481.819(3)  &  6.13(9)  \\
$f2$ & 184.71248(23)  &  5413.819(6)  &  2.63(10) \\
$f3$ & 184.74037(25)  &  5413.002(7)  &  2.32(10)\\
$f4$ & 191.84271(36)  &  5212.604(10)  &  1.63(9) \\
$f5$ & 196.31012(10)  &  5093.981(2)  &  6.44(9)  \\
$f6$ & 198.19774(30)  &  5045.466(8)  &  2.03(9)  \\
$f7$ & 209.14104(20)$^{\dag}$  &  4781.462(45) & 0.35(9) \\
$f8$ & 212.71520(22)$^{\dag}$  &  4701.122(49) & 0.32(9) \\ \hline
\multicolumn{4}{l}{$^{\dag}$ These frequencies are only above the detection}\\
\multicolumn{4}{l}{threshold in the MDM and McDonald data.}\\
\end{tabular}
\end{table}

\section{Discussion}
\subsection{Multiplet constraints on pulsation modes}
A simple test is to examine the data for constant spacings between
frequencies. Such multiplet structure can be used to constrain the
degree, $\ell$, of the spherical harmonics that describe the pulsation
geometry on the stellar surface (Winget et al. 1991). Even a cursory
glance at our observed frequencies indicates that no two splittings
are close to evenly spaced. Combined with a spectroscopic constraint
on the rotation velocity of $v\sin i\leq 10\, {\rm km}\,{\rm s}^{-1}$
(Heber, Reid, \& Werner 2000),
we are left with three possibilities: 1) Rotation is sufficiently slow
that differing $m$ values are degenerate; 2) our line of sight is 
along the pulsation axis, with $\sin i\approx 0$, making only $m=0$
modes observable because of geometric cancellation (Pesnell 1985); or 3)
internal rotation is such that even though the external (spectroscopic)
rotation is slow, rotationally-induced multiplets are widely spaced and
uneven (Kawaler \& Hostler 2005).

\subsection{Amplitude variability}
Since we are relatively confident that our data have resolved the pulsation
spectrum of KPD~2109, we can examine several features of the pulsations
themselves. If sdBV stars are observed over an extended time period,
it is common to detect amplitude variability in many, if not all, of the
pulsation frequencies.
Such variability can occasionally be ascribed to pulsations too
closely spaced to be resolved in any subsets of the data, but often
appear in clearly resolved pulsation spectra where it cannot be ascribed
to mode beating. PG~1605+072 is perhaps the most
complex sdBV star, pulsating in over 55 frequencies. However, it also
has the highest amplitude pulsations and over several
years of observations, the main peaks do not stay at the same amplitudes,
or even at the same frequencies (O'Toole et al. 2002). PG~1336-018 was
observed to have pulsation amplitudes that changed on a daily
basis during a multisite campaign (Kilkenny et al.
2004). Feige~48 and PG~1219+534 both have simple, readily-resolved pulsation
spectra that show clear amplitude variations over the course of months
to years (Reed et al. 2004; Harms, Reed, \& O'Toole 2005).

Figure~\ref{fig04} shows the amplitudes and phases of the 4 frequencies
which are sufficiently resolved and detectable on
a nightly basis. As our campaign covers 32 nights, we can readily detect
small changes in amplitude that over time form trends but which might go
unnoticed in shorter runs. Additionally, we can use this information for
detecting systematic differences in data between observatories. Though there
was one Athens run (night 3) that was very noisy and two MDM runs (nights
28 and 30) that were very short, and thus have large errors, there are no
trends between observatories, indicating that timing and amplitudes were
not adversely impacted by using the different telescopes. 
For KPD~2109, $f1$ and $f5$ clearly appear stable
over our time scale. Neither the amplitude or phase changes substantially.
The amplitude of $f4$ nearly vanishes around day 2 of our run(JD=2453249)
with a peak amplitude of $\approx 3$~mma on day 12. Interestingly, the
corresponding phase varies by about 40\% and looks nearly sinusoidal,
though we do not know why this should be. However, as the phase becomes
less precise around amplitude minimum, it may just be a happenstance
artifact of the data. While the phases for $f6$ appear essentially stable, 
the amplitude clearly increases steadily over the span of our run; more
than doubling in magnitude. As such, KPD~2109 presents itself as an 
interesting case with two very stable modes and two that obviously vary
in amplitude with at least one of these varying significantly in phase
as well. We cannot readily discuss the $f2$ - $f3$ doublet as we cannot
resolve them in individual runs and $f7$ and $f8$ are not detected
sufficiently to examine their amplitudes or phases. Included in 
Fig.~\ref{fig04} are the amplitudes determined by K98 (filled squares) 
and J04 (filled triangles) for the four frequencies shown. Except for
the J04 amplitude for the 5212~$\mu$Hz frequency, they are consistent
with what we find. There are no obvious longer--period trends, though
these would be difficult to find considering the amplitude variability
within our 32 night data set.

\subsection{The nature of the excitation mechanism}
We can also use the criterion outlined in Pereira \& Lopes (2005) to 
examine the nature of the excitation mechanism.
If the frequencies are stochastically excited, we can expect to find
the standard deviation of the amplitudes, $\sigma (A)$, divided by the
average amplitude $\langle A\rangle$ to 
have a value near 0.5 (Eqn. 7 of Pereira \& Lopes 2005). 
Both parameters and their ratio for the frequencies
in Fig.~\ref{fig04} are given in Table~\ref{tab05}. As might be expected
based solely on their amplitudes, $f1$ and $f5$ have very small ratios 
indicative of non-stochastically excited modes while $f6$ almost exactly
fits the criterion for stochastically excited pulsations. Frequency $f4$ 
is slightly ambiguous; appearing near-enough to 0.5 to make its support
of stochastic excitation conceivable, but not convincing.

\begin{table}
\centering
\caption{Standard deviation and mean amplitude for four readily resolvable
frequencies. \label{tab05}}
\begin{tabular}{cccc}
Frequency & $\sigma (A)$ & $\langle A\rangle$ &$\sigma (A)/\langle A\rangle$ \\
 ($\mu$Hz) & (mma) & (mma) & \\ \hline
5481 &  6.20 & 0.45 & 0.07  \\
5212 & 1.68 & 0.72 & 0.43 \\
5093 & 6.49 & 0.75 & 0.12 \\
5045 & 2.37 & 0.12 & 0.51 \\ \hline
\end{tabular}
\end{table}

\subsection{Constraints on pulsation degree via mode density}
Another question involving sdBV stars is the mode degree $\ell$ of the
pulsations. In resolved sdBV stars, we sometimes observe many more 
pulsation modes than $\ell$=0, 1, and 2 can provide. Higher $\ell$ modes
may be needed, but if so they must have a larger intrinsic
amplitude because of the large degree
of geometric cancellation (Charpinet et al. 2005; Reed, Brondel, \& 
Kawaler 2005). From the lack of multiplet structure we may assume that
all $m$ values are degenerate and only concern ourselves with the 
number of $\ell$ modes available within the observed frequency limits
of KPD~2109. Not attempting to match frequencies, but merely choosing
a representative model based on $\log g$ and $T_{\rm eff}$ from 
spectroscopy (Heber et al. 2000), we can determine if a model can provide
the observed mode density using only $\ell$=0, 1, and 2 modes. We searched
our model grid of ISUEVO models (see Reed et al. 2004) within the
spectroscopic error box of KPD~2109. The minimum distance between 
consecutive orders for the same $\ell$ was $\approx 1050\mu$Hz
with a median spacing of $\approx 1200\mu$Hz for $\ell$=0, 1, or 2. We
also examined a representative model from the published grid by 
Charpinet et al. (2002; model \# 7 with $T_{\rm eff}=31311$~K and
$\log g=5.75$). The spacings between their consecutive orders near the
appropriate frequency range for KPD~2109 was 892, 1413, 575, and 949~$\mu$Hz
for $\ell$=0, 1, 2, and 3, respectively. As the observed range between 
$f1$ and $f6$ is only $\approx 500\,\mu$Hz, models can only supply a single
frequency per $\ell$ degree. To obtain the number of frequencies
observed in KPD~2109, assuming no rotational splitting, would require
$\ell$ values up to 5. Thus we can add KPD~2109 to the list of sdBV stars
with pulsations too dense to be accounted for with only low-degree 
($\ell\leq 2$) modes.

\subsection{Comparison with multicolour photometry}
As part of the discovery data, K98 obtained 4 nights (spread over 6 days
and totaling 26.6 hours) of simultaneous 4-color (UBVR)
observations of KPD~2109 and in 2002, three consecutive nights of ULTRACAM 
data (totaling 9.7 hours) were obtained by Jeffery et al. (2004; hereafter
J04). We have frequencies $f1$ - $f6$ in common with K98 who,
using unconstrained and adiabatically constrained phases
identifies $\ell$= 1, 2, 2, 2, 2, 1 and 1, 0, 2, 2, 2, 1 respectively.
We have $f1$, $f2$, $f4$, $f5$, and $f6$ in common with J04 who identifies
these frequencies as $\ell$= 0, 2, 1, 0, and 2. Interestingly, J04 identifies
the 5084~$\mu$Hz frequency detected in K98 as $\ell$=4. However, we again
(and disappointingly) have to dismiss this frequency as undetectable in
the J04 data set. J04 were unable to simultaneously fit all of their data
with their frequency solutions because of the complexity of the data window.
However, for their 3 individual runs, the frequency resolutions are
245, 153, and 40~$\mu$Hz, respectively; far too large to distinguish the
5084~$\mu$Hz frequency from the much higher amplitude 5093~$\mu$Hz one.
The final result is that none of the frequencies have the same
$\ell$ value in all three determinations and none of the reliably detected
frequencies indicate $\ell$ values larger than 2, though such values are
required to match the frequency density given current models.

\section{Conclusions}
Based on the extensive data acquired from five observatories, we are able to
resolve the pulsation spectrum of the pulsating sdB star KPD~2109+4401.
We have analyzed the entire data set as well as several subsets to insure
that we have detected real peaks rather than aliases.
We confirmed the five known 
frequencies $f1$ -- $f5$ previously
identified by both K98 and Bill\'{e}res et al.\ (1998), but do
not detect the 5084~$\mu$Hz frequency detected by K98. We suspect
that as this frequency was the lowest amplitude pulsation detected by
K98, the amplitudes showed some variability in his data,
and since it must have been somewhat masked by the window pattern
(see Fig.~4 of K98), that it was likely an artifact of his data. Of course
another possibility is that the 
amplitude has since diminished beyond our detection limit.  We resolve
a doublet ($f2$ and $f3$) also detected by K98, but at 
0.8~$\mu$Hz separation, was not readily resolvable in his data.
Furthermore, our best data sets
detect the feature at 4781~$\mu$Hz, which was suspected by K98 and 
marginally detect another frequency at 4701~$\mu$Hz. 

As shown in Fig.~\ref{fig02} and provided in Table~\ref{tab03},
all of our subsets detect the same frequencies without ambiguity, though
the doublet is only resolved in Groups I and II.
The detection of the two low-amplitude frequencies
($f8$ and $f9$) relies on the high S/N data from larger
telescopes; here the MDM 1.3-m and the McDonald 2.1-m telescopes. We have
shown that even though there is residual power in the FT after prewhitening
by the six main frequencies, it is almost certainly caused by
amplitude variation.

We have examined our frequency content for observational constraints
on the pulsation modes with the following results, and detect no signs
of evenly spaced frequency multiplets that could be induced by rotation.
As such, we have no way to distinguish between pulsation frequencies of varying
$m$ values and assume that they are degenerate, or $m=0$. Current models
cannot reproduce the observed frequency density without invoking high-degree
($\ell\geq 3$) modes. However, multicolour photometry constrains all
observed frequencies to $\ell\leq 2$, though the identifications for
individual frequencies disagree between methods and observers. 

We examined pulsation amplitudes of the four frequencies that 
are resolvable on a night-by-night basis over the duration of our run
and notice that two frequencies are stable while two 
vary substantially. The amount of amplitude variability can be used
to test if the pulsations are of a stochastic nature and we determine
that two frequencies are definitely not stochastically excited while the
two amplitude-variable frequencies could be.

Our scientific goal of this observational study was to 
resolve the pulsation
structure of the sdB star KPD~2109 by combining 
limited amounts of data from larger telescopes with data from
smaller ($\sim$0.5~m) telescopes.
As expected, this combination has allowed us
a long timebase sufficient to resolve closely spaced pulsations of the star
and the increased S/N of the larger telescopes allow us to detect
pulsations with amplitudes as low as 0.3\,mma. These data are five times better
than K98 in resolution and three times more sensitive in amplitude.
These successful efforts
encourage us to carry out detailed follow-up observations for other
poorly studied stars of this class.

ACKNOWLEDGMENTS: We would like to thank the MDM and McDonald Observatory TACs 
for generous time allocations. This material is based upon work supported 
by the National Science Foundation under Grant No. 0307480.
Any opinions, findings, and conclusions or recommendations expressed 
in this material are those of the author(s) and do not necessarily 
reflect the views of the National Science Foundation.

\end{document}